\documentstyle[eqsecnum,prl,aps,twocolumn]{revtex}
\input epsfig.sty

\begin{document}

\wideabs{
\preprint{HEP/123-qed}
\title{First search for gravitational wave bursts with a network of detectors
}
\author{Z.A. Allen,$^a$ P. Astone,$^b$ L.Baggio,$^c$ D. Busby,$^a$ M. Bassan,$^{d,e}$ 
D.G. Blair,$^f$ M. Bonaldi,$^g$ P. Bonifazi,$^{h,b}$  P. Carelli,$^i$ \\
  M.Cerdonio,$^c$ 
E. Coccia,$^{d,e}$ L. Conti,$^c$ C. Cosmelli,$^{j,b}$ V. Crivelli
Visconti,$^c$ S. D'Antonio,$^k$ 
V. Fafone,$^k$  \\
 P. Falferi,$^g$ P. Fortini,$^l$ S. Frasca,$^{j,b}$ W.O. Hamilton,$^a$ I.S. Heng,$^a$ 
E.N. Ivanov,$^f$ W.W. Johnson,$^a$ \\
 M. Kingham,$^a$ C.R. Locke,$^f$ A. Marini,$^k$ 
V. Martinucci,$^m$ E. Mauceli,$^k$ M.P. McHugh,$^a$ R. Mezzena,$^m$ Y. Minenkov,$^e$ \\
 I. Modena,$^{d,e}$ G. Modestino,$^k$ A. Moleti,$^{d,e}$ A. Ortolan,$^n$ G.V.
Pallottino,$^{j,b}$ G. Pizzella,$^{d,k}$ 
G.A.Prodi,$^{m,*}$\\ 
 E. Rocco,$^c$ F. Ronga,$^k$ F. Salemi,$^j$ G. Santostasi,$^a$
 L. Taffarello,$^o$ 
R. Terenzi,$^{h,e}$ M.E. Tobar,$^f$ G. Vedovato,$^m$\\ 
 A. Vinante,$^m$ M. Visco,$^{h,e}$ S. Vitale,$^m$ 
L. Votano,$^k$ J.P. Zendri,$^o$ \\
(International Gravitational Event Collaboration)
}
\address{
$^a$ Department of Physics and Astronomy, 
Louisiana State University, Baton Rouge, Louisiana 70803}
\address{ 
$^b$ I.N.F.N., Sezione di Roma1, P.le A.Moro 2, I-00185, Roma, Italy}
\address{ 
$^c$ Dipartimento di Fisica, Universit\`a
di Padova, and I.N.F.N., Sezione di Padova, Via Marzolo 8, 35131 Padova,
Italy}
\address{ 
$^d$ Dipartimento di Fisica, Universit\`a 
di Roma ``Tor Vergata'', Via Ricerca Scientifica 1,  I-00133 Roma, Italy}
\address{ 
$^e$ I.N.F.N. Sezione di Roma2, Via Ricerca Scientifica 1,  I-00133 Roma,Italy}
\address{ 
$^f$ Department of Physics,
University of Western Australia, Nedlands, WA 6907
Australia}
\address{ 
$^g$ Centro di Fisica degli Stati Aggregati, I.T.C.- C.N.R., and
I.N.F.N., Trento, I-38050 Povo, Trento, Italy}
\address{ 
$^h$ Istituto Fisica Spazio Interplanetario, 
C.N.R., Via Fosso del Cavaliere, I-00133 Roma, Italy}
\address{ 
$^i$ Dipartimento di Fisica, Universit\`a de L'Aquila, and I.N.F.N., L'Aquila,Italy}
\address{
$^j$ Dipartimento di Fisica, Universit\`a di Roma
``La Sapienza'', P.le A.Moro 2, I-00185, Roma, Italy}
\address{
$^k$ Laboratori Nazionali di Frascati,
Istituto Nazionale di Fisica Nucleare,
Via E.Fermi 40, I-00044, Frascati, Italy}
\address{ 
$^l$ Dipartimento di Fisica, Universit\`a di Ferrara, and I.N.F.N.,
Sezione di Ferrara I-44100 Ferrara, Italy}
\address{
$^m$ Dipartimento di Fisica, Universit\`a
di Trento, and I.N.F.N., Gruppo Collegato di Trento,
I-38050 Povo, Trento, Italy}
\address{ 
$^n$ Laboratori Nazionali di Legnaro,
Istituto Nazionale di Fisica Nucleare, 35020 Legnaro, Padova, Italy}
\address{
$^o$ Sezione di Padova, Istituto Nazionale di Fisica Nucleare, Via
Marzolo 8, I-35131 Padova, Italy
}
\maketitle

\begin{abstract}

We report the initial results from a search for bursts of gravitational
radiation by a network of five cryogenic resonant detectors during 1997
and 1998. This is the first significant search with more than two
detectors observing simultaneously. No gravitational wave burst was
detected. The false alarm rate was lower than 1 per $10^4$ years when
three or more detectors were operating simultaneously. The typical
threshold was $H\simeq 4 \times 10^{-21} Hz^{-1}$ on the Fourier
component at $\sim 10^3\ Hz$ of the gravitational wave strain amplitude.
New upper limits for amplitude and rate of g.w. bursts have been set.

\end{abstract}
\pacs{95.85.Sz, 04.80.Nm, 95.45.+i}
} 


The direct detection of gravitational waves will be a watershed event
for both the physics of gravitation and the investigation of compact
astronomical objects. A variety of astrophysical events are expected to
produce gravitational waves of short duration ($\ll 1$ sec), or {\it gw
bursts}, such as the gravitational collapse of stars or the final few
orbits and the subsequent coalescence of a close binary system of
neutron stars or black holes \cite{sources}. Due to the inherent
weakness of such signals, and the difficulty in distinguishing them from
a myriad of noise sources, the direct detection of a gw burst will
require coincident detection by multiple detectors with uncorrelated
noise. Searches for gw bursts over periods of observation of $1-3$
months have been performed in the past by pairs of cryogenic resonant
bar detectors
\cite{89-lsuromastanford,99-allegroexplorer,99-explorernautilusniobe},
setting upper limits on the incoming rate. A few days of observation
have been reported for simultaneous operation of three cryogenic bar
detectors \cite{89-lsuromastanford} and, with much less sensitivity, of
a pair of short-arm interferometric detectors \cite{96-garchingglasgow}.
Upper limits on gw signals from coalescing binaries has been recently
reported also by a single interferometric detector for $25$ hours of
observation \cite{99-caltechPRL}.

In the last few years, the increase of the number of cryogenic resonant
detectors in simultaneous operation has greatly improved the prospects
of obtaining a confident detection of gw bursts. There are now
five operational cryogenic bar detectors: {\it ALLEGRO} (Baton Rouge,
Louisiana, USA) \cite{AL}, {\it AURIGA} (Legnaro, Italy) \cite{AU}, {\it
EXPLORER} (CERN) \cite{EX}, {\it NAUTILUS} (Frascati, Italy) \cite{NA}
and {\it NIOBE} (Perth, Australia) \cite{NI}. The groups operating these
detectors agreed in $1997$ to start a global search for short ($\sim
1~ms$) gw bursts under common protocols, by establishing the
International Gravitational Event Collaboration ({\it IGEC})
\cite{IGEC}. 

All these detectors use the same principles of operation. The gw excites
the first longitudinal mode of the cylindrical bar, which is cooled to
cryogenic temperatures to reduce the thermal noise and is isolated from
seismic and acoustic disturbances. To measure the strain of the bar, a
secondary mechanical resonator tuned to the cited mode is mounted on one
bar face and a sensor measures the displacement between the secondary
resonator and the bar face. The resulting noise of the detectors
in terms of strain at the input is $5-10 \times
10^{-22}/\sqrt{\mathrm Hz}$ in a bandwidth of $\sim 1\ Hz$ surrounding
the two coupled-mode frequencies. Some of the important physical
parameters of the five detectors are shown in Table~1. The
detector response is optimal for a gw incoming perpendicular to the bar
axis and polarized along it \cite{sin2theta}. The axes of all the bar
detectors are aligned to within a few degrees of one another, so that
the chance of coincidence detection is maximized. This makes the
amplitude acceptance of the detectors for the Galactic Center direction
greater than $0.7$ for about $60\%$ of the time \cite{GWDAW}. 

\begin{figure}[h] \centering
\epsfig{figure=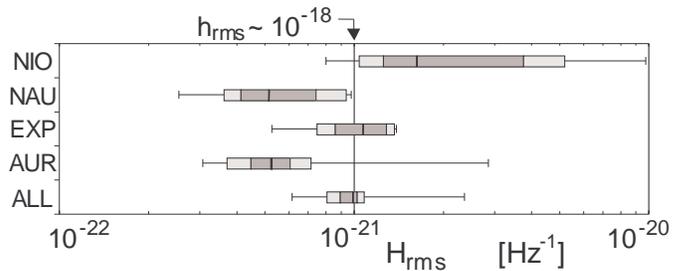} 
\caption{ 
Spread of the noise of detectors during 1997-1998 in terms of the
Fourier component $H_{rms}$ of the gw at $SNR=1$. The plotted
bands of variability of the noise are delimited by selected
percentiles, i.e. by selected fractions of the observation time for
which the sensitivity has been better than $H_{rms}$: bold tick $50\%$,
gray band $16\--84\%$, white band $2.5\--97.5\%$, "T" lines $0\--100\%$.
The corresponding gw amplitude $h_{rms}$ for a $\sim 10^{-3}s$ burst is
sketched in the upper scale.
}
\label{nonstationary} 
\end{figure}

Each detector output is processed by filters optimized for short gw
bursts, giving the estimate for the Fourier component $H(\omega)$ of the
strain amplitude $h(t)$ in the detection bandwidth of $\sim 1\ Hz$
around the mode frequencies listed in Table~1. More specifically, $h(t)$
is the gw amplitude multiplied by the antenna pattern of the detector
\cite{sin2theta}. With the exception of the {\it ALLEGRO} detector, the
noise of the detectors was typically not stationary over long
observation times and was affected by some unmodeled noise sources,
whose correlation with common environmental noise sources was found to
be weak \cite{niobe95}. Fig.~\ref{nonstationary} shows the
variability of the noise of each detector during 1997-1998, in terms of
the Fourier component of the gw corresponding to unity signal-to-noise
ratio, $H_{rms}$. The detectors had quite similar noise levels, since
the typical values of $H_{rms}$ were all within a factor of 3.

We point out that this search for bursts is suitable for any transient
gw which shows a nearly flat Fourier transform $H(\omega)$ of its
amplitude $h(t)$ at the two resonant frequencies of each detector. The
metric perturbation $h(t)$ can either be a millisecond pulse, a signal
made by a few millisecond cycles or a signal sweeping in frequency
through the detector resonances. The IGEC search is therefore sensitive
to different kinds of gw sources such as a stellar gravitational
collapse \cite{sources}, the last stable orbits of an spiralling $NS$
or $BH$ binary, its merging and its final ringdown \cite{flanagan98}.
The computation of $h$ from the measured Fourier component $H$
requires a model for the signal shape. A conventionally chosen shape is
a pulse lasting $\sim 10^{-3}s$ \cite{GR15}; in this case, $H$ should be
multiplied by $\sim 10^3 Hz$ to get the corresponding strain amplitude,
$h$.
 
This letter reports the results of the first coincidence search for gw
bursts performed by the IGEC observatory. The observations covered most
of $1997-1998$, including $625.0$ days with at least one detector in
operation, $260.4$ days with at least two detectors in simultaneous
operation, $89.7$ days with three detectors, and $15.5$ days with four.
This is the first search with significant observation time with more
than two detectors. The duration of simultaneous operation would have
been greater if it had been possible to operate these instruments
with higher duty factors, which were typically $\lesssim 50\%$
during this period with the exception of {\it ALLEGRO}. More details on
the observatory, its data exchange protocol and the exchanged data set
can be found in Ref. \cite{GWDAW}. 

The analysis of the data can be divided into two parts: a generation of
candidate {\it event lists} for each of the individual detectors, and a
time coincidence analysis using the lists. This approach, though
not optimal, has the advantage of being easily implemented and
provides for a satisfactory effectiveness.

Each IGEC group extracted the candidates for gw bursts, or {\it events},
by applying a threshold to the filtered output of the detector. The
events were described by their Fourier magnitude $H$, their arrival
time, the detector noise at that time and other auxiliary information. To
limit the expected rate of accidental coincidences, each detector
threshold was adaptively set to obtain a maximum event rate of $\sim 100$ /
day, with typical values in the range $H_{det}\sim 2 \-- 6 \times 10^{-
21} Hz^{-1}$ corresponding to magnitude signal-to-noise ratio SNR $
\simeq 3 \-- 5$. Single spurious excitations are vetoed against
disturbances detected by environmental sensors. The {\it AURIGA}
detector checked each event against the expected waveform template by
means of a $\chi^2$ test \cite{chi2}. The lists of the events exchanged
within IGEC by each detector also include declarations of the off- and
on- times for the detectors.


All searches for coincident events used a time window of 1.0 second.
This choice limits the false dismissal probability to less than a few
per cent while it ensures a very low false alarm probability when at
least three detectors are observing simultaneously. No three- and
four-fold coincidence was detected, and therefore we did not identify
candidates for gravitational wave detection in the $89.7$ days of
three-fold observation. The detector thresholds were typically $3 \times
10^{-21} Hz^{-1}$ for the most sensitive three-fold configuration ({\it
ALLEGRO-AURIGA-NAUTILUS}) and $5 \times 10^{-21} Hz^{-1}$ for the
others. To give examples of detectable signals, these thresholds
would correspond to respectively $\sim 0.04$ and $0.11\ M_\odot$ converted to
isotropic radiation in the optimal polarization at the distance of
the Galactic Center ($10\ kPc$), assuming a g.w. burst of $1~ms$
duration \cite{GR15}. For comparison, the signal expected from the last
stable orbits of an optimally oriented $NS$ coalescing binary at $10\
kPc$ with $2\times1.4\ M_{\odot}$, would give $H(\omega) = 3\-- 4 \times
10^{- 21} Hz^{-1}$ at the detector resonant frequencies. The number and
amplitude of the two-fold coincidences found in the $260.4$ days of
two-fold observation are in agreement with the estimated accidental
background \cite{GWDAW}.

The estimation of the false alarm rate is a crucial element in any gw
search. It allows for the interpretation of any observed coincidences as
well as the evaluation of the potential of the observatory. Since the
events arrival times of each detector are randomly distributed with a
non stationary rate, the expected background of accidental coincidences
can be computed by two methods: i) by modeling the event times as
Poisson point process and using the measured rates of events for each
detector, and ii) by counting the coincidences after performing even
time shifts of the event times of one detector with respect to the
others \cite{weber}. 

In the first approach,
the expected rate of accidental coincidences is \cite{accidental}
 
\begin{equation}
\lambda = N\frac{(\Delta t)^{N-1}}{T^{N}_{obs}}\prod^{N}_{i=1}
n_{i} ,\label{eq:one}
\end{equation}

\noindent 
where $N$ is the number of detectors simultaneously
operating, $T_{obs}$ their common observation time, $\Delta t = 1\ s$
the maximum time separation for a coincidence, $n_{i}$ the
number of events of the $ith$ detector during $T_{obs}$.
 This equation holds 
even if the event rates of detectors are not stationary as long as
they are uncorrelated among different detectors. 

The second approach is more empirical. In the case of the two-fold
coincidence searches, the time shift results are in agreement with those
predicted through Eq.~\ref{eq:one} \cite{GWDAW}, and demonstrate that
the event rates of different detectors are uncorrelated.

\begin{figure} \centering
\epsfig{figure=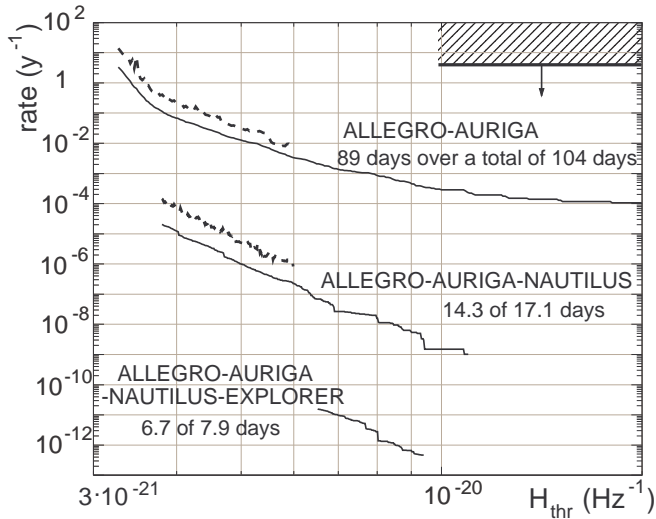}
\caption{
Estimated rate of accidental coincidences, $\lambda \ [year^{-1}]$,
versus the threshold  $H_{thr} \ 
[Hz^{-1}]$ for a sample pair, triple and four-tuple of detectors in 1997-
1998. The continuous lines show the mean value of $\lambda$ for signal
amplitudes $\geq H_{thr}$. 
The dashed lines represent the one $std.\ dev.$ upper bounds for the time variation
of the instantaneous accidental rates. This figure takes into account the
best 85\% of common observation times, when
every detector had an event search threshold lower than 
$3.25$, $3.8$ and $6.5 \ \times 10^{-21} Hz^{-1}$, respectively for the pair,
the triple and the four-tuple. 
The $\lambda$ for the other operative configurations of detectors were similar,
allowing for a small increase of the corresponding $H_{thr}$, at most by a factor
of 2.
The bold horizontal line with arrow stands for the new upper
limit set by all IGEC detectors on the rate of incoming gw
bursts during 1997-1998.
}
\label{accidental}
\end{figure}

The capabilities of the IGEC observatory with respect to the false alarm
probability are shown in Fig.~\ref{accidental} for a few sample
configurations of the observatory. The accidental rate is calculated as
a function of a signal amplitude threshold $H_{thr}$ at the detectors by
applying Eq.~\ref{eq:one} to the number of events of the detectors whose
amplitude is $\geq H_{thr}$. The typical time variability of the
instantaneous accidental rate $\lambda$ has been calculated by means of
a Monte Carlo simulation based on the measured non-stationary behavior
of event rates on single detectors. This variability is about one order
of magnitude with respect to the mean and is mainly determined by the
non stationary performances of the detectors. The estimated mean
background of two-fold coincidences is still fairly high, unless
$H_{thr}$ is raised well above the data exchange threshold $H_{det}$. On
the other hand, a three-fold or four-fold coincidence search keeps a
high statistical significance even for $H_{thr} \sim H_{det}$, since the
expected accidental rates are low enough: respectively less than 1 false
alarm per $10^{4}$ or $10^{6}$ years of observation at $H_{thr} \sim 4
\times 10^{-21} Hz^{-1}$. These fall rapidly as $H_{thr}$ increases. In
fact, the IGEC accidental background noise would remain negligible even
after centuries of observation time.


The $260$ days of observation with two or more detectors in simultaneous
operation improved by about a factor of three the previously set upper
limit on the rate of gw bursts incident on the Earth
\cite{99-allegroexplorer}. Assuming the emission is described by a
stationary Poisson point process and using the same procedure as in Ref.
\cite{89-lsuromastanford,99-allegroexplorer}, the limiting rate set for
a gw burst emission from isotropically distributed sources is
$\leq 4\ year^{-1}$ for $H_{gw} \geq 10^{-20} Hz^{-1}$
(Fig.~\ref{accidental}) with 95\% confidence. A more complete analysis is in
progress.


The IGEC observatory can also be used to set an upper limit on the
amplitude of gw bursts corresponding to astronomical events, such as
supernovae or gamma ray bursts. For time windows of the order of the
hour or larger, each detector is likely to show accidental events and
therefore this upper limit benefits from a multiple coincidence search
among the operating detectors. 

\begin{figure}[b] \centering
\epsfig{figure=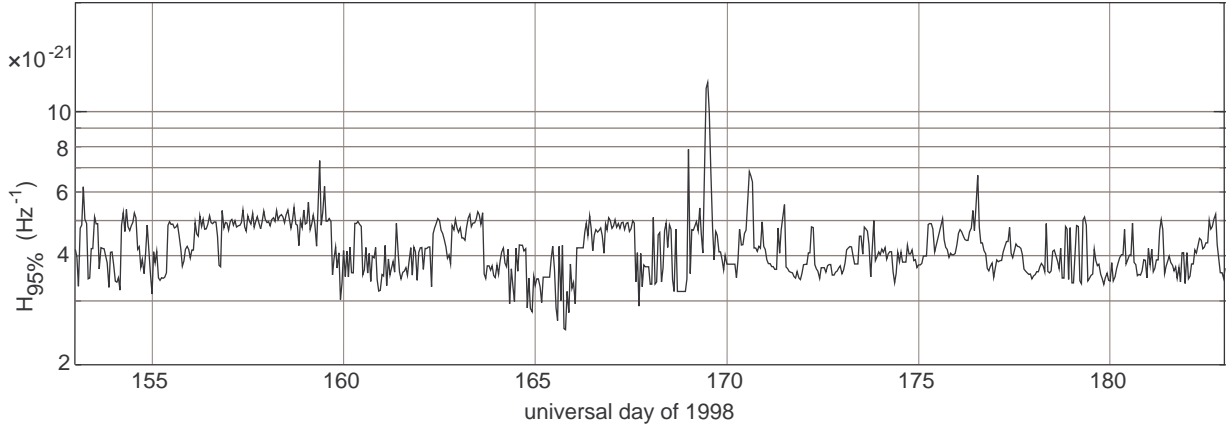}
\caption{
A sample of the upper limit with 95\% confidence on the amplitude of single gw
bursts incident with optimal polarization and orientation on the IGEC observatory
hour by hour in June 1998. The highest peaks shown are due to
single high amplitude events of one detector while the others were not operating.
}
\label{fig:amp}
\end{figure}

A sample of the upper limits on the amplitude of gw bursts occurring
within a time span of $1$ hour, is shown in Fig.~\ref{fig:amp} for a few
weeks of 1998, when up to four detectors were operating. These
limits apply to the component of the radiation emitted with optimal
polarization from a source optimally oriented with respect to the
detectors. We are $95\%$ confident that there was no radiation above
this level hour by hour. 
In all of 1998, the limits set by the IGEC
observatory were better than $H_{gw} = 6$ and $4 \times 10^{-21}
Hz^{-1}$ for $94\%$ and $21\%$ of the year respectively.
To specialize these upper limits for a specific source
direction, each detector response should be divided by its antenna
pattern \cite{sin2theta}. The corresponding observation times of the
Galactic Center by IGEC within the same limits have been
respectively $44\%$ and $7.5\%$ of 1998. For a source at the Galactic
Center emitting isotropically a $1\ ms$ burst \cite{GR15}, the
above upper limits correspond to about $0.16$ and $0.07 \ M_\odot$
converted in the optimal polarization. 

Finally, we remark that the IGEC observatory is capable of monitoring
the strongest galactic sources with a very low false alarm probability
when at least three detectors are simultaneously operating. 
All the
groups involved are actively working for upgrading the current detector
performances and therefore we expect in the near future to extend
the observation range to the Local Group of galaxies, which means an
increase of a factor of 10 of the observed mass.

The ALLEGRO group was supported by the National Science Foundation grant
9970742 and LSU, the NIOBE group by the Australian Research Council, and the Italian
groups by the Istituto Nazionale di Fisica Nucleare and
in part by a grant COFIN'97 from Ministero dell'Universit\`a e della
Ricerca Scientifica e Tecnologica.

\widetext
\begin{table}
\caption{
Main characteristics of the IGEC cryogenic bar detectors.
The detectors measure the mean Fourier component $H$ of the
gw in the detection bandwidth of $\sim 1\ Hz$ around the mode
frequencies. $H = (4L\nu^2)^{-1}\sqrt{E/M}$, where $E$ is the energy deposited in the bar
by the gw and $\nu$ is the mean of the mode frequencies. The bars are
made by Al5056 except for NIOBE, whose bar is made of Nb. The sub-kelvin
detectors and NIOBE showed very similar typical energy sensitivity in
1997-1998, better of a factor of about 4 with respect to the other
detectors. The differences in mass and material, though, affect the gw
sensitivity and give a conversion factor from $\sqrt{E}$ to $H$ which is
2.3 times worse for NIOBE than for the other detectors.
}
\begin{tabular}{lccccc}
detector &ALLEGRO&AURIGA&EXPLORER&NAUTILUS&NIOBE\\
\tableline
Mode frequencies [$Hz$] & 895, 920& 912, 930& 905, 921& 908, 924& 694, 713\\
Bar mass $M$ [$kg$] & 2296& 2230 & 2270 & 2260 & 1500 \\
Bar length $L$ [$m$] & 3.0 & 2.9 & 3.0 & 3.0 & 2.75 \\
Bar temperature [$K$] & 4.2 & 0.2 & 2.6 & 0.1 & 5.0 \\
Longitude & $91^\circ 10' 44" W$ & $11^\circ 56' 54" E$ & $6^\circ 12' E$ & $12^\circ 40' 21" E$ & $115^\circ 49' E$ \\
Latitude & $30^\circ 27' 45" N$ & $45^\circ 21' 12" N$ & $46^\circ 27' N$ & $41^\circ 49' 26" N$ & $31^\circ 56' S$ \\
Azimuth & $40^\circ W$ & $44^\circ E$ & $39^\circ E$ & $44^\circ E$ & $0^\circ$ \\
\end{tabular}
\label{detectors}
\end{table}

\narrowtext

\twocolumn

\widetext


\begin{references}
\bibitem{sin2theta}
a bar detector detects the fraction $sin^2(\theta)\ cos(2\psi)$ of the amplitude of
an incident gw, where $\theta$ is the angle
between the bar axis and the gw direction and $\psi$ is the
polarization angle of the gw in the wavefront plane with respect to the bar axis.
\end{references}
\end{document}